\documentclass[singlespacing]{elsart}

\usepackage{graphicx}

\usepackage{amssymb}
\journal{Vibrational Spectroscopy}
\begin{document}

\begin{frontmatter}

\title{Analytic description of inversion vibrational mode for ammonia molecule}

\author{ A.E. Sitnitsky},
\ead{sitnitsky@mail.kibb.knc.ru}

\address{Kazan Institute of Biochemistry and Biophysics, P.O.B. 30, Kazan
420111, Russia. Tel. 8-843-231-90-37. e-mail: sitnitsky@kibb.knc.ru }

\begin{abstract}
The one-dimensional Schr\"odinger equation with symmetric trigonometric double-well potential is exactly solved via angular prolate spheroidal function. Although it is inferior compared with multidimensional counterparts and its limitations are obvious nevertheless its solution is shown to be analytic rather than commonly used numerical or approximate semiclassical (WKB) one. This comprises the novelty and the merit of the present work. Our exact analytic description of the ground state splitting can well be a referee point for comparison of the accuracy of numerous WKB formulas suggested in the literature. The approach reasonably well suits for the inversion mode in the ammonia molecule $NH_3$ and thus yields a new theoretical tool for its description. The results obtained provide good quantitative description of relevant experimental data on microwave and IR spectroscopy of $NH_3$.
\end{abstract}

\end{frontmatter}

\section{Introduction}
Ammonia molecule $NH_3$ is a very important object both in itself and for the development of physics. In particular it played a cornerstone role in the development of radiospectroscopy and quantum electronics as the basis for the first maser \cite{Tow55}. Up to now it remains a subject of intensive researches (see \cite{Sut93}, \cite{Han99}, \cite{Raj04}, \cite{Sal10}, \cite{Pol16}, \cite{Hug00}, \cite{Jan14}, \cite{Spi27} and refs. therein). Besides from the very beginning of quantum mechanics it is a "work horse" object for verification of analytic and numerical methods of treating Schr\"odinger equation (SE) with a double-well potential \cite{Hun27}, \cite{Ros32}, \cite{Man35}, \cite{Sut93}. For ammonia molecule a pertinent degree of freedom (inversion mode) in the vibrational spectrum of $NH_3$ corresponds to the motion of the $N$ atom relative to the $H_3$ symmetrical triangle \cite{Tow55} (physically three hydrogen atoms undergo simultaneous tunneling from one side of the nitrogen atom to the other \cite{Cro05} although the nitrogen also slightly moves to provide that the position of the center of mass remains constant). As a result a fictitious quantum particle with the reduced mass
\begin{equation}
\label{eq1} M=\frac{3m_Hm_N}{3m_H+m_N}
\end{equation}
moves under the influence of a double-well potential (DWP) along the coordinate corresponding to the distance of the $N$ atom from the $H_3$ plane \cite{Tow55}. DWP results from the Coulomb repulsion between the nitrogen nucleus and the three protons. This picture is somewhat simplified and more complex paths for atoms movement at inversion are suggested \cite{Tow55}. However these subtleties are not important for the present analysis. We remain within the framework of an approach in which $NH_3$ inversion description is reduced to the solution of one-dimensional SE with a pertinent double well potential and parameters chosen to suit corresponding experimental data of microwave and IR spectroscopy. We fully aware that our SE is inferior compared with multidimensional counterparts and its limitations are obvious. Nevertheless its solution is shown to be exact analytic rather than commonly used numerical or approximate semiclassical (WKB) one. This comprises the novelty and the merit of the present work.

SE with a DWP finds many well-known applications in physics and chemistry beginning from the above noted inversion of $NH_3$ and ending by heterostructures, Bose-Einstein condensates and superconducting circuits (see \cite{Xie12}, \cite{Dow13}, \cite{Che13}, \cite{Har14}, \cite{Dow16}, \cite{Dow17}, \cite{Tur10}, \cite{Tur16} and refs. therein). We mention here only smooth double-well potentials and leave aside numerous models with rectangular wells or sewing together two single-well potentials (harmonic, Morse, etc.) that pervade textbooks and pedagogical-style articles on quantum mechanics.
There is noticeable advance in obtaining quasi-exact (i.e., exact for some particular choice of potential parameters) \cite{Xie12}, \cite{Dow13}, \cite{Che13} and exact (those for an arbitrary set of potential parameters) \cite{Har14}, \cite{Sit17} solutions for such SE by their reducing to the confluent Heun's equation (CHE). A lot of potentials for SE are shown to be solvable via the confluent Heun's function (CHF) \cite{Ish16}. The CHF is a well described special function tabulated in {\sl {Maple}} \cite{Fiz12}, \cite{Fiz10}, \cite{Sha12}. The latter makes its usage to be a routine procedure. This fact renders the obtained solution of SE to be very convenient for applications. Recently the exact solution of the Smoluchowski equation for reorientational motion in Maier-Saupe DWP was obtained via CHF \cite{Sit15}, \cite{Sit16}. The method yields the probability distribution function in the form convenient for application to nuclear spin-lattice relaxation \cite{Sit11}. Here we apply similar approach to SE with a practically important type of DWP.

In the present article we modify the trigonometric DWP suggested in \cite{Sit17} for the description of intrinsically asymmetric hydrogen bonds to that including the symmetric case as a particular one. The trigonometric DWP is a particular case of some general potential from \cite{Ish16} (N2 with $m_{1,2}=\left(1/2,1/2\right)$ from Table.1). However the specific form of the potential investigated in \cite{Sit17} can not be directly applied to the symmetric case. Let us denote the dimensionless coordinate as $y$. The form in \cite{Sit17}
\[
 U(y)=h\ \tan^2 y+\sqrt h \frac{\sin y}{\cos^2 y}-b\sin^2 y+a\sin y
\]
is intrinsically asymmetric. Even if we make the depths of the wells to be equal by an appropriate choice of $a$ there is still an asymmetry in barrier shape. Here we consider the form
\begin{equation}
\label{eq2} U(y)=h\ \tan^2 y-b\sin^2 y+a\sin y
\end{equation}
This potential is truly symmetric at $a=0$.

The aim of the article is to solve exactly SE with the latter potential, to determine the quantization rule for obtaining energy levels and to provide analytic representation of the wave functions for these energy levels. We show that the wave functions are expressed via Coulomb (generalized) spheroidal function (CSF). In the particular case of symmetric potential ($a=0$) the latter is reduced to an ordinary spheroidal function that is realized in {\sl {Mathematica}}. We apply our results to the inversion of the ammonia molecule $NH_3$ that is a classical example of the process with a symmetric DWP.

The physics of $NH_3$ (and more generally of tunneling phenomena including energy levels splitting) is very well understood by now as a result of qualitative arguments, semiclassical (WKB) analysis and numerical solution of SE with DWP. The stringent analytic results for the exactly solvable trigonometric DWP are complementary to this understanding. The results obtained agree well with previous knowledge and supplement it with precise calculations within the framework of the particular potential (trigonometric DWP). Also it should be noted that in the present article we do not touch upon the problem of the rotation-vibration interaction, i.e.,  we assume $NH_3$ molecule to be on the lowest $J=1; K=1$ rotational level (the ground state of para-ammonia \cite{Cro05}).

The paper is organized as follows.  In Sec. 2 the problem under study is formulated.  In Sec. 3 and 4 the solution of SE is presented via CHF and CSF respectively. In Sec. 5 the case of symmetric potential is treated.
In Sec.6 the results for $NH_3$ are discussed and the conclusions are summarized. In Appendix 1 and Appendix 2 some minor technical details are presented.

\section{Schr\"odinger equation and the potential}
We consider the one-dimensional SE
\begin{equation}
\label{eq3} \frac{d^2 \psi (x)}{dx^2}+\frac{2M}{\hbar^2}\left[E-U(x)\right]\psi (x)=0
\end{equation}
where $U(x)$ is a DWP that is infinite at the boundaries of the finite interval $x=\pm L$.
We introduce the dimensionless energy $\epsilon$ and the dimensionless distance $y$
\begin{equation}
\label{eq4} \epsilon=\frac{8ML^2E}{\hbar^2 \pi^2}\ \ \ \ \ \ \ \ \ \ \ \ \ \ \ \ \ \ \ \ \ \ \ \ \ \ \ \ \ \ \ \ y=\frac{\pi x}{2L}
\end{equation}
so that $-\pi/2\leq y \leq \pi/2$. The dimensionless SE with the potential (\ref{eq2}) takes the form
\begin{equation}
\label{eq5} \psi''_{yy} (y)+\left[\epsilon-h\ \tan^2 y+b\sin^2 y-a\sin y\right]\psi (y)=0
\end{equation}
where $h$ is the barrier width parameter, $b$ is the barrier height parameter and $a$ is the asymmetry parameter (see Appendix). An example of this potential is depicted in Fig.1.
\begin{figure}
\begin{center}
\includegraphics* [width=\textwidth] {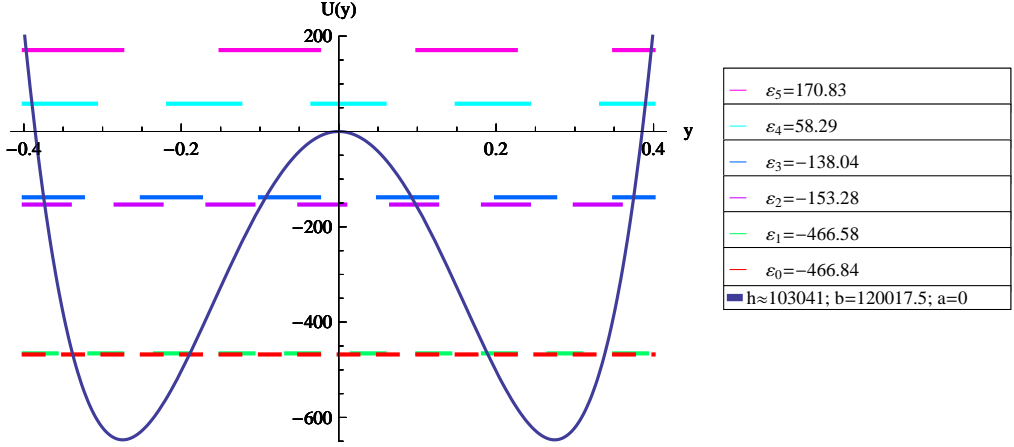}
\end{center}
\caption{The model double-well potential (\ref{eq2}) at the values of the parameters $m=\sqrt{h+1/4}=321$ ($h\approx 103041$), $b=120017.49$, $a=0$. The parameters are chosen to describe the potential and the energy levels for ammonia molecule
$NH_3$ (experimental data are taken from \cite{Hug00}, \cite{Her45}). The energy levels $\epsilon_0=-466.84$, $\epsilon_1=-466.58$, $\epsilon_2=-153.28$, $\epsilon_3=-138.04$, $\epsilon_4=58.29$, $\epsilon_5=170.83$ are respectively depicted by the dashes of increasing length.} \label{Fig.1}
\end{figure}

\section{Solution of the Schr\"odinger equation via the confluent Heun's function}
We introduce a new function $\varphi (y)$ by the relationship
\begin{equation}
\label{eq6} \psi (y)=\cos^{1/2} y\ \exp \left(-\sqrt {b}\sin y \right)\left[ \tan\left(\frac{\pi}{4}+\frac{y}{2}\right)\right]^{\sqrt {h+1/4}}\varphi (y)
\end{equation}
and a new variable
\begin{equation}
\label{eq7} z=\frac{1+\sin y}{2}
\end{equation}
The equation for $\varphi (z)$ is
\[
z(z-1)\varphi''_{zz}(z)+\left[-4\sqrt{b} z^2+2\left(2\sqrt{b}+1\right)z-\left(\sqrt{h+1/4}+1\right)\right]\varphi'_z(z)+
\]
\begin{equation}
\label{eq8}
\left[2\left(a-2\sqrt{b}\right)z+2\sqrt{b}\left(\sqrt{h+1/4}+1\right)+1/4-\epsilon-h-b-a\right]\varphi (z)=0
\end{equation}
It belongs to the type of CHE. For the infinite at the boundaries potential the wave function must be zero there. As a result the solution of our SE consistent with this requirement has the form
\[
\psi (y)=\cos^{1/2} y\ \exp \left(-\sqrt {b}\sin y \right)\left[ \tan\left(\frac{\pi}{4}+\frac{y}{2}\right)\right]^{\sqrt {h+1/4}}\times
\]
\begin{equation}
\label{eq9} {\rm HeunC}\Biggl(-4\sqrt {b},\sqrt {h+1/4},-\sqrt {h+1/4},2a,
\frac{3}{8}-a-b-\frac{h}{2}-\epsilon;\frac{1+\sin y}{2}\Biggr)
\end{equation}
At $y=-\pi/2$ the wave function is automatically zero due to $\cos$ and $\tan$. However at $y=\pi/2$ the function $\tan$ diverges and $\cos$ can not cope with this disaster at $h>0$ without the help from the side of ${\rm HeunC}$.
The above mentioned requirement can be satisfied only by a specific constraint imposed on ${\rm HeunC}$ and thus yields the boundary condition for determining the energy levels. As a result we obtain the equation for eigenvalues by setting $y=\pi/2$
\begin{equation}
\label{eq10}{\rm HeunC}\Biggl(-4\sqrt {b},\sqrt {h+1/4},-\sqrt {h+1/4},2a,
\frac{3}{8}-a-b-\frac{h}{2}-\epsilon;1\Biggr)=0
\end{equation}
Its solutions form the spectrum of eigenvalues $\epsilon_n$ where $n=0,1,2, ...\ $ for the energy $\epsilon$.

{\sl {Maple}} enables one to solve easily and efficiently (\ref{eq10}) and to plot the wave function (\ref{eq9}). However even the normalization of the wave function (let alone the calculation of matrix elements with its help) requires dealing with integrals containing {\rm HeunC}. Unfortunately {\sl {Maple}} carries out this task extremely inefficiently (see \cite{Fiz12} where some drawbacks of ${\rm HeunC}$ realization in {\sl {Maple}} are expertly discussed). Attempts to circumvent this difficulty lead to tiresome numerical calculations that practically cancel out all the advantages of our analytic solution compared with a numerical solution of SE. For this reason we further look for another form of the solution that gives much more convenient computational tool for the case of symmetric potential. We remind that namely the latter is necessary for the description of inversion mode in $NH_3$.

\section{Solution of the Schr\"odinger equation via the Coulomb spheroidal function}
We introduce a new variable
\begin{equation}
\label{eq11} s=2z-1
\end{equation}
where $-1\leq s \leq 1$ and a new function $v(s)$ by the relationship
\begin{equation}
\label{eq12} \varphi (s)=\left(\frac{1-s}{1+s}\right)^{\frac{\sqrt{h+1/4}}{2}}\exp\left(s\sqrt {b}\right)v(s)
\end{equation}
The equation for $v(s)$ is obtained from (\ref{eq8})
\[
\frac{d}{ds}\left[\left(1-s^2\right)\frac{dv(s)}{ds}\right]+
\]
\begin{equation}
\label{eq13} \left[b+\epsilon+h-\frac{1}{4}-\sqrt {b\left (h+\frac{1}{4}\right)}-b\left(1-s^2\right)-as-\frac{h+1/4}{1-s^2}\right]v(s)=0
\end{equation}
We denote
\begin{equation}
\label{eq14} h=m^2-\frac{1}{4}
\end{equation}
If $m$ is integer then (\ref{eq13}) belongs to the type of Coulomb (generalized) spheroidal equations \cite{Kom76}
and its solution is
\begin{equation}
\label{eq15} v(s)=\bar\Xi_{mq}\left(\sqrt {b}, -a;s\right)
\end{equation}
where $q=0,1,2,...$ and $\bar\Xi_{mq}\left(\sqrt {b}, -a;s\right)$ is CSF. The energy levels are determined from the relationship
\begin{equation}
\label{eq16} \epsilon_q=\lambda_{mq}\left(\sqrt {b}, -a\right)+\frac{1}{2}-b-m^2
\end{equation}
where $\lambda_{mq}\left(\sqrt {b}, -a\right)$ is the spectrum of eigenvalues for $\bar\Xi_{mq}\left(\sqrt {b}, -a;s\right)$. CSF is normalized by the requirement \cite{Kom76}
\begin{equation}
\label{eq17} \int_{-1}^1ds\  \bar\Xi_{mq'}\left(\sqrt {b}, -a;s\right)\bar\Xi_{mq}\left(\sqrt {b}, -a;s\right)=\delta_{qq'}
\end{equation}

As a result the wave function takes the form
\[
\psi_q (y)=\cos^{1/2} y\ \left[ \tan\left(\frac{\pi}{4}+\frac{y}{2}\right)\right]^m\left(1-\sin y\right)^{\frac{m}{2}}\times
\]
\begin{equation}
\label{eq18} \left(1+\sin y\right)^{-\frac{m}{2}}\bar\Xi_{mq}\left(\sqrt {b}, -a;\sin y\right)
\end{equation}
The formal proof that the product of the Coulomb spheroidal function appearing in (\ref{eq18}) with the divergent trigonometric functions is zero at $y=\pi/2$ is given in Appendix 2.
Unfortunately the CSF is realized neither in {\sl {Maple}} nor in {\sl {Mathematica}} at present and the solution (\ref{eq18}) is practically inapplicable for calculations.

\section{Symmetric potential}
For the symmetric potential ($a=0$) CSF is reduced to an angular prolate spheroidal function \cite{Kom76}
\begin{equation}
\label{eq19} \bar \Xi_{mq}\left(\sqrt{b}, 0;s\right)=\bar S_{m(q+m)}\left(\sqrt{b};s\right)
\end{equation}
where $m$ is integer and $q=0,1,2,...\ $. The latter is realized in  {\sl {Mathematica}} as $\bar S_{m(q+m)}\left(\sqrt{b};s\right)\equiv SpheroidalPS[(q+m),m,i\sqrt{b},s] $ and is normalized by the requirement \cite{Kom76}
\begin{equation}
\label{eq20} \int_{-1}^1ds\ \bar S_{m(q+m)}^2\left(\sqrt{b};s\right)=1
\end{equation}

Thus we have in this case
\[
\psi_q (y)=\cos^{1/2} y\ \left[ \tan\left(\frac{\pi}{4}+\frac{y}{2}\right)\right]^m\left(1-\sin y\right)^{\frac{m}{2}}\times
\]
\begin{equation}
\label{eq21} \left(1+\sin y\right)^{-\frac{m}{2}}\bar S_{m(q+m)}\left(\sqrt{b};\sin y\right)
\end{equation}
The formal proof that the product of the angular prolate spheroidal function appearing in (\ref{eq21}) with the divergent trigonometric functions is zero at $y=\pi/2$ is given in Appendix 2.
The energy levels are determined from the relationship
\begin{equation}
\label{eq22} \epsilon_q=\lambda_{m(q+m)}\left(\sqrt {b}\right)+\frac{1}{2}-b-m^2
\end{equation}
where $\lambda_{m(q+m)}\left(\sqrt {b}\right)$ is the spectrum of eigenvalues for $\bar S_{m(q+m)}\left(\sqrt{b};s\right)$. It is realized in {\sl {Mathematica}} as $\lambda_{m(q+m)}\left(\sqrt {b}\right)\equiv SpheroidalEigenvalue[(q+m),m,i\sqrt {b}]$.

The formulas (\ref{eq21}) and (\ref{eq22}) provide a highly efficient and convenient tool for calculating the wave functions and the energy levels of SE with symmetric trigonometric potential (\ref{eq2}) with the help of {\sl {Mathematica}}.

\section{Results and discussion}
Fig.1 shows that the parameters of the potential (\ref{eq2}) can be chosen to provide good description of the energy levels structure for a set of specific experimental data. In Fig.1 the energy levels for the inversion of the ammonia molecule $NH_3$ (experimental data determined with help of microwave and IR spectroscopy are taken from \cite{Hug00}, \cite{Her45}) are presented. The energy levels for the inversion mode of $NH_3$ form a pair of doublets within the wells. For the ground-state splitting the values $E_1-E_0=0.66\div0.8\ {\rm cm^{-1}}$ are given in the literature \cite{Her45}, \cite{Hug00}, \cite{Jan14}. To be specific we choose the value $E_1-E_0=0.76\ {\rm cm^{-1}}$. For the transition frequency of the upper state $E_2-E_0$ there seems to be a consensus $E_2-E_0=932.4\ {\rm cm^{-1}}$ \cite{Her45}, \cite{Hug00}, \cite{Jan14}. However for the transition frequency of the upper state $E_3-E_0$ also a range of values exists $E_3-E_0=968.4\div 977.3\ {\rm cm^{-1}}$ ($E_3-E_2=35.7\div 45.1\ {\rm cm^{-1}}$) \cite{Her45}, \cite{Hug00}, \cite{Jan14}. We obtain the best description by taking the value $E_3-E_0=977.3\ {\rm cm^{-1}}$ ($E_3-E_2=45.1\ {\rm cm^{-1}}$) \cite{Hug00}.  The distance between the minima of the potential $2l\approx 0.76\ \AA$ is known from the geometry of $NH_3$ \cite{Vol49}, \cite{Her45}. These experimental values are obtained from our dimensionless ones if we take $m=321$ ($h\approx 103041$), $b=120017.49$, $a=0$ (with taking into account that for hydrogen $m_H=1\ {\rm amu}$ and for nitrogen $m_N=14\ {\rm amu}$ so that (\ref{eq1}) yields $M\approx 2.47\ {\rm amu}$). From (\ref{eq4}) we have the relationship of the position for the dimensional minimum $l\approx 0.38\ \AA$ \cite{Vol49}, \cite{Her45} with our dimensionless one $y_{min}$
\begin{equation}
\label{eq23} y_{min}=\frac{l}{2\hbar}\sqrt \frac{8M\left( E_{1}-E_{0}\right)}{\epsilon_{1}-\epsilon_{0}}
\end{equation}
From our value $y_{min}\approx 0.27$ it follows that for the conversion coefficient we should set the value $L\approx 2.23 \ \AA$.

The normalized wave functions are obtained from (\ref{eq21}) as follows
\begin{equation}
\label{eq24} \psi_i(y)^{normalized}=\psi_i(y)\left\{\int_{-\pi/2}^{\pi/2}dy\ \left(\psi_i(y)\right)^2\right\}^{-1/2}
\end{equation}

They are depicted in Fig.2. All these functions are highly delocalized as it should be for a symmetric DWP. The functions $\psi_0(y)^{normalized}$ and $\psi_1(y)^{normalized}$ correspond to the split ground state. The ground state splitting is very small ($0.76\ {\rm cm^{-1}}$ in dimensional units). The functions $\psi_0(y)^{normalized}$ and $\psi_2(y)^{normalized}$
are even while those $\psi_1(y)^{normalized}$ and $\psi_3(y)^{normalized}$ are odd. The functions $\psi_2(y)^{normalized}$ and $\psi_3(y)^{normalized}$ describe excited states. It should be stressed that the latter are rather close to the barrier top (see Fig.1). Thus the functions $\psi_2(y)^{normalized}$ and $\psi_3(y)^{normalized}$ can not be described in semiclassical approximation (WKB). The exact analytic representation of these wave functions is the merit of the present approach. In contrast the functions of the split ground state can be treated by WKB method (see \cite{Tur10}, \cite{Tur16} for review).
In recent years there is noticeable progress in developing this method (see \cite{Gar00}, \cite{Son08}, \cite{Ras12}, \cite{Son15} and refs. therein) but different variants of ordinary semiclassical approximation along with instanton approach are suggested in the literature. Our exact analytic result can well be a referee point for comparison of the accuracy of suggested formulas.

\begin{figure}
\begin{center}
\includegraphics* [height=5cm] {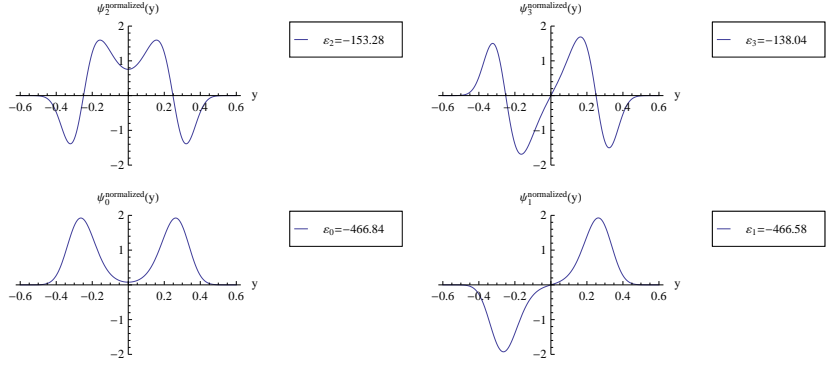}
\end{center}
\caption{Normalized wave functions (\ref{eq21}) for Schr\"odinger equation (\ref{eq5}) with the double-well potential (\ref{eq2}) corresponding to the energy levels $\epsilon_0=-466.84$, $\epsilon_1=-466.58$, $\epsilon_2=-153.28$, $\epsilon_3=-138.04$. The parameters of the potential are chosen to describe the energy levels for the inversion vibrational mode in ammonia molecule $NH_3$ (experimental data are taken from \cite{Hug00}, \cite{Her45}). The splitting of the ground state $\epsilon_1-\epsilon_0=0.26$ corresponds to $0.76\ {\rm cm^{-1}}$ in dimensional units.} \label{Fig.2}
\end{figure}

Both CSF and ordinary spheroidal functions are defined for integer $m$ \cite{Kom76}. As a result we have to limit ourselves by a discrete set of the values for the parameter $h=m^2-1/4$  (\ref{eq14}). For these reason we do not pretend to provide a very precise description of $NH_3$ with the help of spheroidal functions. The application of CHF (making use of (\ref{eq9}) and (\ref{eq10})) circumvents this difficulty (i.e., the approach is valid for arbitrary values of $h$) and hence is able to provide more accurate description. However as was mentioned above the calculations of integrals containing CHF lead to tiresome numerical manipulations that deprive our approach of its advantage compared with a numerical solution of SE. In contrast {\sl {Mathematica}} tackles the integrals containing spheroidal functions very efficiently. Thus we conclude that the formulas (\ref{eq21}) and (\ref{eq22}) are more convenient for calculations and provide reasonably good accuracy.

In the present article we speak about $NH_3$ to be specific. It is obvious however that our approach is applicable to $ND_3$ or $NT_3$ with corresponding definition of the reduced mass in (\ref{eq1}). Moreover  our approach remains valid for any specific problem where one has to deal with a one-dimensional DWP in SE. Three parameters in the trigonometric DWP (\ref{eq2}) enable one to model experimental data of IR spectroscopy or results of quantum chemical calculations of the potential surface for such problem with reasonable accuracy.

One can conclude that the Schr\"odinger equation with symmetric trigonometric double-well potential can be exactly solved via angular prolate spheroidal function. Our exact analytic description of the ground state splitting can well be a referee point for comparison of the accuracy of numerous WKB formulas suggested in the literature. The approach reasonably well suits for the inversion mode in the ammonia molecule $NH_3$. Thus it yields a new theoretical tool for the description of this important molecule. The results obtained provide reasonably good quantitative description of relevant experimental data on microwave and IR spectroscopy of $NH_3$.

\section{Appendix 1}
For the symmetric DWP ($a=0$) the parameters $h$ and $b$ can be related with the barrier height $B=-U\left(y_{min}\right)$ and barrier width $\Delta=y_{min}^{(1)}-y_{min}^{(2)}$ as follows
\[
\Delta=2\arccos\left(\frac{h}{b}\right)^{1/4}
\]
\[
B=\left(\sqrt {h}-\sqrt {b}\right)^2
\]
Inversely one obtains
\[
b=\frac{B}{\left\{1-\left[\cos\left(\Delta/2\right)\right]^2\right\}^2}
\]
\[
h=\frac{B\left[\cos\left(\Delta/2\right)\right]^4}{\left\{1-\left[\cos\left(\Delta/2\right)\right]^2\right\}^2}
\]

\section{Appendix 2}
Here we provide the formal proof that the product of both the Coulomb and the angular prolate spheroidal functions appearing in (\ref{eq18}) and (\ref{eq21}) respectively with the divergent trigonometric functions is zero at $y=\pi/2$. For both $\bar\Xi_{mq}\left(\sqrt {b}, -a;\sin y\right)$ and $\bar S_{m(q+m)}\left(\sqrt{b};\sin y\right)$ the point $y=\pi/2$ is a regular singular one. In this regular point they behave as \cite{Kom76}
\[
\bar\Xi_{mq}\left(\sqrt {b}, -a;\sin y\right)\sim \left(1-\sin^2 y\right)^{m/2}
\]
\[
\bar S_{m(q+m)}\left(\sqrt{b};\sin y\right)\sim \left(1-\sin^2 y\right)^{m/2}
\]
As a result it is enough to prove that the following limit is zero
\[
\lim_{y\to\pi/2}{\cos^{1/2} y\ \left[ \tan\left(\frac{\pi}{4}+\frac{y}{2}\right)\right]^m \left[\frac{\left(1-\sin y\right)}{\left(1+\sin y\right)}\right]^{\frac{m}{2}}\left(1-\sin^2 y\right)^{m/2}}=0
\]
Making use of the substitution $y=\pi/2-\epsilon$ we obtain after straightforward calculation
\[
\lim_{\epsilon\to 0}{\epsilon^{1/2+m}}=0
\]
for any positive $m$ that proves the above assertion.

The behavior of both (\ref{eq18}) and (\ref{eq21}) results from the requirement that for the infinite at the boundaries potential the wave function must be zero there
\[
\psi_q (y)
\left| {\begin{array}{l}
  \\
{boundary}\\
 \end{array}}\right. =0
\]
This boundary condition does not contain the derivatives of the wave function and hence is of the so-called Dirichlet type. The so-called Neumann boundary condition (containing only the derivatives) or Cauchy one (that of mixed type) can arise in a number of physical systems \cite{Xie12}, \cite{Dow13}, \cite{Che13}, \cite{Har14}, \cite{Dow16}, \cite{Dow17} where a hyperbolic DWP tends to zero at the boundaries which are at $\pm\infty$. For instance the well known Manning potential \cite{Man35} (which was applied to the ammonia molecule long ago) is of that type. It was shown to be a particular case of the general hyperbolic DWP considered in \cite{Har14}. However for $NH_3$ the atoms are bound by covalent bonds and as a result (unless the latter are broken) our fictitious quantum particle discussed in the Introduction can not go to infinity. This situation is physically described by a DWP which is infinite at the finite boundaries. The trigonometric DWP considered in the present article is just the one of the required type.\\

As was mentioned in the text the formulas (\ref{eq21}) with spheroidal function and (\ref{eq22}) with its spectrum of eigenvalues very efficiently provide calculations by means of {\sl {Mathematica}}. The formula (\ref{eq22}) yields the result for the energy of a $q$-th level immediately (at a click). In contrast the usage of CHF requires numerical solution of (\ref{eq10}) with the help of {\sl {Maple}}. Although the latter treats CHF efficiently the search of the required root involves tedious manipulations as compared with the automated call of (\ref{eq22}) in {\sl {Mathematica}}. The difference in computational time in that event is insignificant and the advantage of the present approach compared with previous one is only a matter of convenience. However this is not the case for calculation of integrals containing the wave functions. The usage of (\ref{eq21}) provides considerable advantage in this regard. As was discussed in \cite{Sit17} {\sl {Maple}} is practically unable to calculate integrals with wave functions containing CHF. To circumvent this difficulty one has to resort to tedious manipulations with the series representation of CHF involving the problem of its proper truncation and subsequent verification of required accuracy. In contrast {\sl {Mathematica}} treats the integrals with (\ref{eq21}) directly and efficiently. For instance the calculation of a normalization integral in (\ref{eq24}) proceeds within several tens of seconds with the help of (\ref{eq21}) on a PC of low to average power and this is achieved by direct call of the automated sequence  SpheroidalPS in {\sl {Mathematica}}. One can achieve comparable computational time with the help of (\ref{eq9})
but this includes a lot of preliminary programming for the series representation of CHF with all mentioned above problems. One can conclude that the usage of spheroidal function is not merely a matter of convenience but provides a qualitatively higher level of computational facilities. Unfortunately it has been gained by present only for the symmetric case of trigonometric DWP. However there is no doubt that CSF will be realized in a mathematical software package sooner or later and then the formulas (\ref{eq16}) and (\ref{eq18}) will provide the same level of facilities for the asymmetric case.

Acknowledgements. The author is grateful to Dr. Yu.F. Zuev
for helpful discussions. The work was supported by the grant from RFBR N 15-29-01239.

\end{document}